\documentstyle[12pt,cite]{article}
\newcommand {\beq}{\begin{equation}}
\newcommand {\eeq}{\end{equation}}
\newcommand {\beqa}{\begin{eqnarray}}
\newcommand {\eeqa}{\end{eqnarray}}
\newcommand {\beqan}{\begin{eqnarray*}}
\newcommand {\eeqan}{\end{eqnarray*}}

\newcommand {\Romannumeral}[1]{\uppercase\expandafter{\romannumeral#1}}

\def\rref#1{(\ref{#1})}

\begin{document}
\setlength{\oddsidemargin}{0cm}
\setlength{\baselineskip}{7mm}

\begin{titlepage}
 \renewcommand{\thefootnote}{\fnsymbol{footnote}}
    \begin{normalsize}
     \begin{flushright}
                     TIT-HEP-287\\
                     YITP/U-95-10\\
                    March 1995~~
     \end{flushright}
    \end{normalsize}
    \begin{Large}
       \vspace{1cm}
       \begin{center}
         {\LARGE Scaling Behavior of Ricci Curvature at Short Distance
                 near Two Dimensions} \\
       \end{center}
    \end{Large}

  \vspace{10mm}

\begin{center}
           Yoshihisa K{\sc itazawa}$^{1)}$\footnote
           {E-mail address : kitazawa@phys.titech.ac.jp}{\sc and}
        	Masao N{\sc inomiya}$^{2)}$\footnote
		{E-mail address : ninomiya@yisun1.yukawa.kyoto-u.ac.jp}\\
      \vspace{1cm}
        $^{1)}$ {\it Department of Physics,
Tokyo Institute of Technology,} \\
            {\it Oh-okayama, Meguro-ku, Tokyo 152, Japan}\\
	$^{2)}$ {\it Uji Research Center, Yukawa Institute for Theoretical
Physics,}\\
	   {\it Kyoto University, Uji 611, Japan}\\
\vspace{15mm}

\end{center}
\begin{abstract}
\noindent
We study the renormalization of the Ricci curvature as an example of
generally covariant operators in quantum gravity near two dimensions.
We find that it scales with a definite scaling
dimension at short distance.
The Ricci curvature singularity at the big
bang can be viewed as such a scaling phenomenon.
The problem of the spacetime singularity may be resolved by the scale
invariance of the spacetime at short distance.

\vspace{20mm}
\noindent
PACS numbers : 04.60.+n,~98.80Hw

\end{abstract}

\end{titlepage}
\vfil\eject

Quantum gravity in $2+\epsilon$ dimensions
is capable to describe classical spacetime at long
distance and simultaneously a consistent theory at short distance
\cite{2+epsilon,KN,KKN1}.
Therefore it may provide us insights into the short distance structure
of our universe.
At the short distance fixed point of the renormalization group,
the theory becomes scale invariant. Hence the
spacetime itself is expected to become scale invariant at short distance.
It is very interesting to study the structure of such a scale invariant
spacetime.

Recall that the conformal transformation is part of the
reparametrization group. Therefore the theory with the reparametrization
invariance
always possesses the conformal invariance if we regard the conformal mode
as a matter field. The rest of the reparametrization group is the volume
preserving diffeomorphism.
We quantize the conformal mode just like a matter field. In such a quantization
scheme, the conformal invariance is a crucial symmetry to respect.
However the conformal invariance is known to be broken at the one loop level.
In order to resolve this dilemma, we have adopted the tree action which only
respects the volume preserving diffeomorphism. Our strategy is to recover the
conformal invariance after including the quantum corrections (counter terms).
By doing so, the conformal anomaly can be canceled between the tree
and the loop contributions\cite{KKN2,AKKN}.

It has been shown that this idea is valid at the one loop level
and higher order corrections can also be computed\cite{AKKN,AKNT}.
The tree level action in $D=2+\epsilon$ dimensions at the short distance
fixed point in the renormalization group is
\beq
{\mu ^\epsilon/G^*}\,\int d^D x \sqrt{\hat{g}}\{\tilde{R}L(\psi
,\varphi _i )
- - {1/2}\,\tilde{g}^{\mu\nu}\partial _{\mu} \psi\partial _{\nu} \psi
+{1/2}\,\tilde{g}^{\mu\nu}\partial _{\mu} \varphi _i\partial _{\nu}
\varphi _i\} ,
\label{action1}
\eeq
where $L=1+ {\epsilon/\{{8(D-1)}\}}\,(\psi ^2 - \varphi _i^2)$
and
$G^* = {24\pi\epsilon/{(25-c)}}$.
$\mu$ is a renormalization scale.
In this action $\tilde{g}_{\mu\nu}
=\hat{g}_{\mu\rho}{(e^h)^{\rho}} _{\nu}$ represents the metric
without the conformal mode $\psi$.
Therefore ${h^{\mu}} _{\nu}$ is traceless ${h^{\mu}} _{\mu} =0$.
$\hat{g}_{\mu\nu}$ is a background
metric we have introduced for convenience.
Tensor indices are always raised and lowered by the background metric.
$\tilde{R}$ is the scalar curvature made out of $\tilde{g}_{\mu\nu}$.
$\varphi _i$ denotes $c$ copies of conformally coupled scalar (matter) fields.
If $L = {\epsilon/\{{8(D-1)}\}}\,(\psi ^2 - \varphi _i^2)$,
the action \rref{action1} is nothing
but the classical Einstein gravity.
We have modified the Einstein action minimally
to obtain
a consistent quantum theory.

The fixed point action possesses the global $Z_2$ symmetry: $\psi \rightarrow
- - \psi$. If the vacuum expectation value of $\psi$ field is nonvanishing,
this symmetry is spontaneously broken. Since $\psi$ represents the scale
factor of the universe, the vacuum expectation value of $\psi$ is certainly
nonvanishing in our universe (in fact it is still growing).
In the latter part of this paper, we demonstrate this point by investigating
the classical solution \rref{claeqn}.
We therefore
identify $\psi$ as the order parameter in quantum gravity. In the weak coupling
phase, the vacuum expectation value of $\psi$ is nonvanishing and the $Z_2$
invariance is spontaneously broken. Our universe belongs to this phase.
In the strong coupling phase, the vacuum expectation value of $\psi$ field
vanishes and
the global $Z_2$ invariance is respected. The short distance fixed point
represents
the phase transition point between the two.

In order to study the structure of spacetime at short distance,
we need to renormalize various operators in the theory.
The renormalization of the gauge invariant operators has been
studied up to now in two dimensional quantum gravity
\cite{Polyakov,DDK} and in $2+\epsilon$ dimensions
\cite{KN,KKN1,KKN2,AKKN}.
These operators cannot depend on a
particular coordinate and are always integrated over the
spacetime manifold.
The relevant and irrelevant operators are found to acquire large anomalous
dimensions of $O(1)$. On the other
the anomalous dimensions of the marginal operators are small
and of $O(\epsilon )$ at the fixed point.

However we also would like to renormalize generally covariant
operators in order to
study the structure of spacetime at short distance.
The Ricci curvature $R_{\mu\nu}$ is such an example.
We would like to know the fate of the spacetime singularity
at the beginning of the universe.
We may be able to gain insights into such a question by
studying the scaling behavior of this operator at short distance.
The advantage to consider
this operator is that it is
invariant under the global conformal transformation.
The Riemann curvature $R^{\mu}_{~\nu\rho\sigma}$,
the Einstein tensor and the
energy momentum tensor of matter fields are also
of this type.
Therefore the anomalous dimension of these operators are
expected to be of $O(\epsilon)$.
This fact
simplifies the renormalization procedure of these operators
just like that of the marginal gauge invariant operators.

It has been suggested that we may be able to renormalize
generally covariant
operators if we relax the general covariance to
the covariance under
the volume preserving
diffeomorphism. We have expected the scaling behavior of these operators
\cite{AKKN}.
In this paper we show that these expectations are indeed realized
at the one loop level.

In quantum theory, the Ricci curvature with the canonical dimension
two will acquire the anomalous dimension.
Therefore the conformal mode dependence of this operator
will be renormalized. We parametrize the metric as $g_{\mu\nu} =
\tilde{g}_{\mu\nu} e^{-\phi}$. We separate the conformal mode
dependence of $R_{\mu\nu}$ explicitly
by the conformal transformation.
We reparametrize the conformal mode in terms of $\psi$ which has
the canonical kinetic term in the action \rref{action1}.
Classically they are related as:
\begin{displaymath}
\exp (-\epsilon /4 \,\phi ) = 1+\sqrt{\epsilon/\{{8(D-1)}\}}\,\psi.
\end{displaymath}
In this parametrization, the Ricci curvature becomes
\begin{eqnarray}
R_{\mu\nu} &=& \tilde{R}_{\mu\nu} \nonumber \\
&-& {1/2}\, \tilde{g}_{\mu\nu} \tilde{g}^{\alpha\beta}
\nabla _{\alpha}\nabla _{\beta}(-\sqrt{2/\epsilon}\, \psi
+{1/4}\,\psi ^2 )  \nonumber \\
&-& {1/2}\,\partial _{\mu}\psi \partial _{\nu} \psi
+ {1/2}\, \tilde{g}_{\mu\nu} \tilde{g}^{\alpha\beta}
\partial _{\alpha}\psi\partial _{\beta}\psi + O(\sqrt{\epsilon})
\label{riccic}
\end{eqnarray}
where $\tilde{R}_{\mu\nu}$ is the Ricci curvature made out of
$\tilde{g}_{\mu\nu}$. The covariant derivative $\nabla _{\mu}$
is taken with respect to $\tilde{g}_{\mu\nu}$.

We have decomposed a generally covariant operator
into the collection of the
operators which are covariant only with respect to the
volume preserving diffeomorphism.
These operators need to be subtracted to define renormalized
finite operators. As it turns out that the Ricci curvature
as a whole cannot be multiplicatively renormalized.
In order to circumvent this problem,
we introduce
different coefficients for different operators
and postulate that the renormalized Ricci curvature
to be
\begin{eqnarray}
R_{\mu\nu} &=&
A \tilde{g}_{\mu\nu} \tilde{g}^{\alpha\beta}
\nabla _{\alpha}\nabla _{\beta}\psi
- - B \tilde{g}_{\mu\nu} \tilde{g}^{\alpha\beta}
\nabla _{\alpha}\nabla _{\beta}\psi ^2 \nonumber \\
&+& C \tilde{R}_{\mu\nu} - D \tilde{g}_{\mu\nu} \tilde{R} \nonumber \\
&-& E \partial _{\mu}\psi \partial _{\nu} \psi
+ F \tilde{g}_{\mu\nu} \tilde{g}^{\alpha\beta}
\partial _{\alpha}\psi\partial _{\beta}\psi
\nonumber \\
&+& E^m \partial _{\mu}\varphi _i \partial _{\nu} \varphi _i
- - F^m \tilde{g}_{\mu\nu} \tilde{g}^{\alpha\beta}
\partial _{\alpha}\varphi _i\partial _{\beta}\varphi _i  .
\label{ricciq}
\end{eqnarray}
In this expression, we have ignored subleading operators
in the
$\epsilon$ expansion.
However we have included the operators
which are absent in \rref{riccic}
for the sake of the renormalizability.

As it is shown in the following,
the subtractions which are required to make the operators
finite close
among the operators which are covariant under the volume
preserving diffeomorphism.
Therefore the Ricci curvature can be renormalized in the
form which
preserves the covariance under the volume preserving diffeomorphism.
The coefficients ($A \sim F^m$) of the operators can be determined by
requiring that the renormalized operator \rref{ricciq} should coincide
the classical
Ricci curvature \rref{riccic} at the weak coupling limit, namely
at long distance. Since we can derive the renormalization group
equations
for these coefficients, such calculations can be done
at least numerically. However we are most interested in the
anomalous dimension of the Ricci curvature at short
distance. For such a purpose, it is enough to renormalize
the operator around the short distance fixed point
and we report the results of such a calculation.

The one loop renormalization of the operators are performed
in a standard background gauge we have utilized
in our previous papers.
We can ignore the external ghost fields in the background gauge
at the one loop level.
Firstly we consider the renormalization of the operator
$A \tilde{g}_{\mu\nu} \tilde{g}^{\alpha\beta}\nabla _{\alpha}\nabla
_{\beta}\psi$.
The relevant vertices to renormalize this operator comes only from
$\int d^D x \sqrt{\hat{g}}
{1/2}\,\tilde{g}^{\mu\nu}\partial _{\mu} \psi\partial _{\nu} \psi$
term in the action.
We need to consider two one loop diagrams with an external $\psi$
line, to which the operator is inserted.

It is found that this operator can be renormalized multiplicatively
at the one loop level. The finite operator with the
counter term is
$$  A_0 \tilde{g}_{\mu\nu} \tilde{g}^{\alpha\beta}
 \nabla _{\alpha}\nabla _{\beta}\psi
= A (\tilde{g}_{\mu\nu} \tilde{g}^{\alpha\beta}
\nabla _{\alpha}\nabla _{\beta}\psi)_r ,
$$
where
$A_0 = (1+g/{(2\pi\epsilon)})$ is a bare coefficient and
$(\tilde{g}_{\mu\nu}
\tilde{g}^{\alpha\beta}
\nabla _{\alpha}\nabla _{\beta}\psi)_r$ denotes a renormalized
operator.

By demanding the bare coefficient to be independent of the renormalization
scale, we find a renormalization group equation for the coefficient $A$:
\beq
\mu {dA/{d\mu}}\, = -{G/{(2\pi)}}\, A .
\eeq
Therefore $A \sim \mu ^{-{G/{2\pi}}}$ and hence it vanishes at
short distance where $\mu \rightarrow \infty$.
As it turns out, the renormalization of
$B \tilde{g}_{\mu\nu} \tilde{g}^{\alpha\beta}
\nabla _{\alpha}\nabla _{\beta}\psi ^2$ is just analogous
and the renormalization group
equation for $B$ is
\beq
\mu {dB/{d\mu}}\, = -{G/{(2\pi)}}\, B .
\eeq
Therefore the coefficient $B$ also vanishes at short distance.
We suspect that the vanishing of $A$ at short distance may be
due to the recovery of $Z_2$ symmetry which has been noted
in the fixed point action.

Next we move on to renormalize the remaining operators.
For this purpose, we need to expand the Ricci curvature
$\tilde{R}_{\mu\nu}$
around the background metric
in terms of $h_{\mu\nu}$ field up to the quadratic terms.
The quadratic terms are
\begin{eqnarray}
& & {1/4}\, \nabla _{\mu}h_{\alpha\beta}
\nabla _{\nu}h^{\alpha\beta}
- - {1/2}\, \nabla _{\beta}h_{\alpha\mu}
\nabla ^{\beta}{h^{\alpha}~_{\nu}}
+ {1/2}\, \nabla ^{\beta}h_{\alpha\mu}
\nabla ^{\alpha}h_{\beta\nu}
\nonumber \\
&+& {1/2}\,
\nabla _{\beta}(h^{\alpha\beta}\nabla _{\nu}h_{\alpha\mu})
+ {1/2}\, \nabla _{\beta}(h^{\alpha\beta}\nabla _{\mu}h_{\alpha\nu})
- -{1/2}\,\nabla _{\beta}(h^{\alpha\beta}\nabla _{\alpha}h_{\mu\nu})
\nonumber \\
&-&{1/4}\, \nabla _{\alpha}\nabla _{\nu}(h^2)^{\alpha}~_{\mu}
- -{1/4}\,\nabla _{\alpha}\nabla _{\mu}(h^2)^{\alpha}~_{\nu}
+{1/4}\, \nabla ^{\alpha}\nabla _{\alpha}{(h^2)_{\mu\nu}} .
\label{ricci}
\end{eqnarray}
In this expression, the covariant derivatives are taken with respect to the
background metric.
There are 13 relevant diagrams for the one loop renormalization.
The diagrams constructed by using the first three terms of (6) are
rather involved to evaluate.
In order to simplify this calculation,
we utilize the Riemann's normal coordinates\cite{BB}.
For simplicity the origin of the Riemann's normal
coordinate is taken to be that of the
propagator.
We adopt the doubling trick and complexify $h_{\mu\nu}$
field to further simplify the one loop calculation.
The propagator of $h_{\mu\nu}$ field is
expanded in terms of the curvature at the origin:
\begin{eqnarray}
<h^{\alpha\beta}(p)h_{\mu\nu}(-p)> & = &
G_0~^{\alpha\beta}~_{,\mu\nu} \nonumber \\
& + & {1/p^{2}}\,
(-{1/3}\,\hat{R} + {2/3}\, \hat{R}^{\rho\sigma}
{p_{\rho}p_{\sigma}/p^{2}}\,)
G_0~^{\alpha\beta}~_{,\mu\nu} \nonumber \\
& + & {1/p^{2}}\, (2\hat{R}^{\alpha}~_{\rho}~^{\beta}~_{\sigma}
+{2/D}\,\delta ^{\alpha\beta}\hat{R}_{\rho\sigma})
G_0~^{\rho\sigma}~_{,\mu\nu} +\cdots ,
\label{prop}
\end{eqnarray}
where $G_0~^{\alpha\beta}~_{,\mu\nu} = {1/p^{2}}\,
(\delta ^{\alpha}~_{\mu} \delta ^{\beta}~_{\nu}
+\delta ^{\alpha}~_{\nu} \delta ^{\beta}~_{\mu}
- -{2/D}\, \delta ^{\alpha\beta}\delta _{\mu\nu})$
is the propagator in the flat
spacetime. The second line in the above expression is analogous
to that of a free scalar field.
With this propagator, the total derivatives
of \rref{ricci} give the one loop divergence
of $-{G/{(2\pi\epsilon)}}\,\tilde{R}_{\mu\nu}$.

In this way, we find the finite operators to be
\begin{eqnarray}
& & C_0 \tilde{R}_{\mu\nu} - D_0 \tilde{g}_{\mu\nu} \tilde{R}
 \nonumber \\
& & - E_0 \partial _{\mu}\psi \partial _{\nu} \psi
+ F_0 \tilde{g}_{\mu\nu} \tilde{g}^{\alpha\beta}
\partial _{\alpha}\psi\partial _{\beta}\psi
\nonumber \\
& & + E^m_0 \partial _{\mu}\varphi _i \partial _{\nu} \varphi _i
- - - F^m_0 \tilde{g}_{\mu\nu} \tilde{g}^{\alpha\beta}
\partial _{\alpha}\varphi _i\partial _{\beta}\varphi _i
\nonumber \\
& = & C (\tilde{R}_{\mu\nu})_r - D (\tilde{g}_{\mu\nu} R)_r \nonumber \\
& & - E (\partial _{\mu}\psi \partial _{\nu} \psi)_r
+ F (\tilde{g}_{\mu\nu} \tilde{g}^{\alpha\beta}
\partial _{\alpha}\psi\partial _{\beta}\psi)_r
\nonumber \\
& & + E^m (\partial _{\mu}\varphi _i \partial _{\nu} \varphi _i)_r
- - F^m (\tilde{g}_{\mu\nu} \tilde{g}^{\alpha\beta}
\partial _{\alpha}\varphi _i\partial _{\beta}\varphi _i)_r .
\end{eqnarray}
The bare coefficients are found to be
\begin{eqnarray}
C_0 &=& C(1-{G/{(6\pi\epsilon)}}\,) +D{4G/{(3\pi\epsilon)}}\,
+ E{G/{(12\pi\epsilon)}}\, + E^m{cG/{(12\pi\epsilon)}}\, ,
\nonumber \\
D_0 &=& D(1+{7G/{(6\pi\epsilon)}}\,)
- -C{G/{(3\pi\epsilon)}}\, +E{G/{(24\pi\epsilon)}}\,
+E^m{cG/{(24\pi\epsilon)}}\, ,
\nonumber \\
E_0 &=& E -C{G/{(4\pi\epsilon)}}\, ,
\nonumber \\
F_0 &=& F(1+{G/{(2\pi\epsilon)}}\,) - E{G/{(4\pi\epsilon)}}\,
- -C{G/{(8\pi\epsilon)}}\,  ,
\nonumber \\
E^m_0 &=& E^m -C{G/{(4\pi\epsilon)}}\, ,
\nonumber \\
F^m_0 &=& F^m(1+{G/{(2\pi\epsilon)}}\,) - E^m{G/{(4\pi\epsilon)}}\,
- -C{G/{(8\pi\epsilon)}}\,  .
\label{reneqnrev}
\end{eqnarray}

The renormalization group equations follow and
from these equations we easily find that
\begin{eqnarray}
\mu{d/{d\mu}}\,(C-2D) &=& - {G/{(2\pi)}}\, (C-2D) ,
\nonumber \\
\mu{d/{d\mu}}\,(E-2F) &=& -{G/{(2\pi)}}\, (E-2F) ,
\nonumber \\
\mu{d/{d\mu}}\,(E^m-2F^m) &=& -{G/{(2\pi)}}\, (E^m-2F^m) .
\label{reneqn2}
\end{eqnarray}

Therefore at short distance, we can put $C=2D,E=2F$ and $E^m=2F^m$
in \rref{reneqnrev} which simplifies the renormalization group equations as
\begin{eqnarray}
\mu{d/{d\mu}}\,C &=& -{G/{(2\pi)}}\, C
- - {G/{(12\pi)}}\, E - {cG/{(12\pi)}}\, E^m ,\nonumber \\
\mu{d/{d\mu}}\,E &=& {G/{(4\pi)}}\, C , \nonumber \\
\mu{d/{d\mu}}\,E^m &=& {G/{(4\pi)}}\, C .
\end{eqnarray}
The solution of these equations
which is symmetric between the conformal mode and matter
fields ($E=E^m$)
is found to be
\begin{eqnarray}
C &=& \lambda _{+} \mu^{-\lambda _{+}}
+ \lambda _{-} \mu^{-\lambda _{-}} ,
\nonumber \\
E &=& -{G/{(4\pi)}}\,(\mu^{-\lambda _{+}}
+ \mu^{-\lambda _{-}}) ,
\end{eqnarray}
where $\lambda_{\pm} = {G/{(4\pi)}}\, (1
\pm \sqrt{(2-c)/3})$.
We can show that this is the only nontrivial solution.

The difficulty of this expression is that it becomes
complex for $c>2$.
However we recall that there is a redundant operator in this channel.
Namely the equation of motion with respect $h_{\mu\nu}$
has the same tensor structure with the operator in question.
It is a traceless
symmetric tensor. Therefore we adopt an interpretation that
the operator with the scaling dimension $\lambda _{\pm}$ is redundant.

With this interpretation, the physical renormalization group
equations are rather \rref{reneqn2}.
These coefficients  show the same
scaling behavior with $A$ and $B$. It is also clear that
there is no mixing between
the physical operators.
However they all possess the same scaling exponents.
The Einstein tensor does not acquire the anomalous
dimension after discarding the redundant operator.
It is also the case with the matter energy momentum
tensor which is related to the Einstein tensor by
the equation of motion.
Since the Einstein tensor vanishes identically
in two dimensions, these results are reasonable.

Therefore we conclude that the renormalized Ricci curvature scales
with the anomalous dimension
$-{G/{(2\pi)}}\, = - {12\epsilon/{(25-c)}}
$ at short distance.
The renormalization effect is found to soften the short
distance singularity.

In order to draw physical implications from the
renormalized Ricci curvature let us consider the Robertson-Walker spacetime
in $D$ dimensions with the line element:
\begin{eqnarray}
ds^2 &=& -dt^2 + r(t)^2  \tilde{g}_{ij} dx^i dx^j ,\nonumber \\
\tilde{g}_{ij} &=& \delta _{ij} +{kx_ix_j/{(1-k|\vec x |^2)}}\, .
\end{eqnarray}
The Ricci curvature is
\begin{eqnarray}
R_{tt} &=& (D-1){\ddot r/r}\, ,\nonumber \\
R_{ij} &=& -(r\ddot r +\epsilon (\dot r )^2 +\epsilon k )\tilde g _{ij} ,
\end{eqnarray}
where $\dot r = {dr/dt}$.

The cosmological solution of \rref{action1}
which can be interpreted as the Robertson-Walker
spacetime with $k=0$ is
\beq
\psi ={1/\sqrt{c}}\, \varphi _i = x_0 ,~
h_{\mu\nu} = 0 .
\label{claeqn}
\eeq
In the classical limit the conformal factor is
$\exp({-\phi})\sim {x_0}^{4/\epsilon}$.
By equating the line elements
\begin{eqnarray}
ds^2 &=& \exp ({-\phi})(-dx_0^2 + \tilde{g}_{ij}dx^idx^j) \nonumber \\
&=&-dt^2 + r^2\tilde{g}_{ij}dx^idx^j ,
\end{eqnarray}
we find
$t \sim {x_0}^{D/\epsilon}$
and $r \sim t^{2/D}$\cite{AKKN}.

Using these relations, we find that
the Ricci curvatures $R_{ij}$ and $R_{00}=({dt/{dx_0}}\,)^2R_{tt}$
scales as
${x_0}^{-2}$ in agreement with the canonical dimension.
If we extrapolate these classical scaling behaviors
at the beginning of the universe where $x_0 \sim 0$
, we encounter the divergence
of the Ricci curvature.

Field theoretically we have found that the Ricci curvature
scales at short distance. However the scaling exponents have acquired the
quantum corrections since the gravitational interaction
becomes important at short distance.
The main conclusion of this paper is that the scaling dimension
of the Ricci curvature has
changed from the classical value of $2$ to
\beq
2 -  {12\epsilon/{(25-c)}}
\eeq
to the leading order
in the $\epsilon$ expansion at short distance.
Although the operator becomes softer in quantum theory,
the Ricci curvature still diverges at short
distance. We argue that this is not
a problem since the theory becomes scale invariant
at short distance. In a scale invariant theory, it is natural
to expect scaling behaviors of the operators.
The Ricci curvature singularity at the big
bang can be viewed as a such scaling phenomenon.

Furthermore the quantum gravity in $2+\epsilon$ dimensions is well defined
at the short distance limit.
In fact we have found that the theory can be extended beyond
the big bang and the universe is found to bounce back from the big crunch
\cite{AKKN}.
In quantum gravity near two dimensions,
we conclude that the
problem of the spacetime
singularity is resolved by the scale invariance of
the spacetime at short distance.
In fact it is a critical phenomena with a universal
scaling exponent which is calculable in the $2+\epsilon$
dimensional expansion.
Our investigation also suggests the
similar mechanism to resolve the spacetime
singularity problem at the big bang in our universe.

We are grateful for discussions on this subject
to our collaborators,
especially T. Aida and H. Kawai.
This work is supported in part by the Grant-in-Aid for Scientific Research
from the Ministry of Education,Science and Culture.

\newpage
\setlength{\baselineskip}{7mm}

\end{document}